\documentclass[english,aps]{revtex4}
\usepackage{setspace}

\makeatletter

\providecommand{\tabularnewline}{\\}

\usepackage{setspace}

\usepackage{babel}
\makeatother
\begin{document}

\title{KDTREE 2: Fortran 95 and C++ software to efficiently search for near
neighbors in a multi-dimensional Euclidean space}

\author{Matthew B. Kennel}

\email{mkennel@ucsd.edu}


\affiliation{University of California, San Diego}

\begin{abstract}
Many data-based statistical algorithms require that one find
\textit{near or nearest neighbors} to a given vector among a set of
points in that vector space, usually with Euclidean topology. The k-d
data structure and search algorithms are the generalization of
classical binary search trees to higher dimensional spaces, so that
one may locate near neighbors to an example vector in $O(\log N)$ time
instead of the brute-force $O(N)$ time, with $N$ being the size of the
data base. KDTREE2 is a Fortran 95 module, and a parallel set of C++
classes which implement tree construction and search routines to find
either a set of $m$ nearest neighbors to an example, or all the
neighbors within some Euclidean distance $r.$ The two versions are
independent and function fully on their own. Considerable care has
been taken in the implementation of the search methods, resulting in
substantially higher computational efficiency (up to an order of
magnitude faster) than the author's previous Internet-distributed
version. Architectural improvements include rearrangement for memory
cache-friendly performance, heap-based priority queues for large
$m$searches, and more effective pruning of search paths by geometrical
constraints to avoid wasted effort.  The improvements are the most
potent in the more difficult and slowest cases: larger data base
sizes, higher dimensionality manifolds containing the data set, and
larger numbers of neighbors to search for. The C++ implementation
requires the Standard Template Library as well as the BOOST C++
library be installed.  {\tt www.arxiv.org e-print: physics/0408067}
Open source software is available at
\url{ftp://lyapunov.ucsd.edu/pub/nonlinear/kd_tree/kdtree2.[tgz|zip]}.
\end{abstract}
\maketitle

\section{introduction}

Given a fixed data base of points on the real line, how would one
efficiently find the closest point to some example $q$ (the query),
assuming that the question will be asked for many different arbitrary
$x$ values? The solution here is classical and obvious: sort the
original data base, and perform a recursive binary bisection for each
query, successively narrowing the range of the original data points
which are possible neighbors to the query point, until the true nearest
neighbors have been found. On average it will take $O(\log N)$ bisections
to locate the nearest neighbor, much less than the effort needed to
\emph{exhaustively} search all $N$ points and remember the one with
the closest distance to $q$. Although for an algorithm this simple
it is not usually implemented this way, one may view the binary search
as progressing down a tree of depth $c\,\log N$, with each node in
the tree specifying a interval, namely the support of the points that
it represents. Descendents of any node partition the interval represented
by their parent, without overlap. Searching for nearest neighbors
involves descending the particular nodes of the tree whose support
intervals contain the query point.

The k-d tree is the natural generalization to a $k$-dimensional Euclidean
space, instead of the 1-d line. Each node of the k-d tree now has
an associated support hyper-rectangle (outer product of $k$ intervals)
instead of a simple 1-d interval. As before each non-terminal node
has two descendants, whose hyper-rectangles partition the parent's
support space in two, along a certain dimension known as the \textit{cut
dimension,} which is chosen in a data-dependent way by the tree building
algorithm. Similarly, each node is associated with a subset of the
data base elements; this subset is guaranteed to lie within each node's
associated hyper-rectangle, known as the \textit{bounding box.} The
non-trivial algorithmic complication, compared to searches of a line,
is that one may need to search for near neighbor candidates contained
not only in those nodes whose bounding boxes contain the query (even
though one generally expects the best matches there), but through
a number of additional neighboring nodes. The trick is to minimize
the number of these {}``extra'' searches which must be performed.

K-d trees are clever, but not magic. The infamous {}``curse of
dimensionality'' still reigns, and will effectively thwart even good
k-d tree methods when the underlying dimensionality of the point set
(i.e. the dimension of a manifold containing the database) is
sufficiently high. Why?  In high-dimensional spaces, as opposed to our
intuition trained in 2 or 3 dimensional space, the distance to even
near or nearest neighbors is almost as large as the distance to a
random point. In practice this means that for sufficiently high
dimension, the k-d tree search methods end up having to search many
{}``extra'' nodes and offer little improvement over brute-force
exhaustive searching. K-d trees are excellent for 3 dimensional data,
offer significant benefits up to perhaps 10-15 dimensional data, but
are useless for 100 dimensional data.

\section{Build and search algorithms}

K-d trees have been discussed a number of times before in the literature;
we refer the reader to \cite{Moore91} for a good tutorial and the
references therein. I will discuss the specific implementation choices
in the KDTREE2 package. Firstly, the KDTREE2 package implements ONLY
the Euclidean metric, i.e. finding near neighbors relative to the
squared distance between points $x$ and $y$, $d^{2}(x,y)=\sum_{i=1}^{d}(x_{i}-y_{i})^{2}$.
The usual alternative situation is for periodic topology in some coordinates.
That can be simulated easily by converting each periodic component
$\theta$ of the original space into two components in Euclidean space,
$(A\cos\theta,A\sin\theta)$ with some scaling factor $A$ as appropriate.

\subsection{Building}

The software assumes a fixed, unchanging input data base of points,
and builds a tree structure for the entire data set. Each tree node
retains local data specifying the range of points (as viewed through
a separately maintained index) represented by the node, the identity
and value of which dimension the children ({}``cut dimension'' and
{}``cut value'') are split on, pointers to the left and right child
nodes, and the bounding box for all points in the node. The root node
is constructed corresponding to all points, and the bounding box explicitly
computed with exhaustive enumeration of the maximum and minimum values
for each coordinate.

Given a bounding box, the build algorithm splits upon that dimension
with the maximum extent, i.e. the largest difference between maximum
and minimum values. The split location is defined (initially) as the
arithmetic average of the values in that coordinate, and then using
an efficient one-pass algorithm, the indexes are moved appropriately
for the two children, corresponding to points with coordinate less
than (left child) and greater than (right child) the cut value.

During the top-down build process, only an approximate bounding box
is used for efficiency. Finding the max and min of points along each
dimension, especially when the database is large, can be time consuming,
requiring $O(N\cdot d)$ work for each level of the tree. The approximation
is to recompute the bounding box only for that dimension that the
parent split upon, and otherwise copy the parent's bounding box for
all other dimensions. This creates bounding boxes which are overestimates
of the true bounding box. When the number of points in any node is
less than or equal to a user-specified constant known as the \emph{bucket
size}, no children are attached, and this becomes a terminal node
of the k-d tree.

Once the points have been identified with nodes, the exact bounding
boxes are then refined in an efficient process taking logarithmic
time. The true bounding boxes are computed explicitly for the terminal
nodes. Once these are created, the exact bounding boxes for any internal
node can be computed rapidly as unions of its children's bounding
boxes. As a result, using approximate bounding boxes in the tree creation
has little effect on search performance, but significantly reduces
the time necessary to build the tree. Along the way, the cut value
is refined to the arithmetic average between the maximum of the box
on the left node and the minimum of box on the right node.

Once the complete tree has been created, the build procedure optionally
can create a new internal copy of the data set, permuting its order
so that data for points in the terminal nodes lie contiguously in
memory, and near points for nodes close topologically. This is recommended
for performance when the database is too large for the main CPU cache.
In that circumstance, the rearrangement approximately doubles performance.

\subsection{Searching}

The search algorithm is simple, but implemented carefully for efficiency.
There are two search modes, fixed neighbor number search (find the
closest $m$ points to the query vector $q$), and the fixed radius
search (find all points $x_{i}$ with $d^{2}(x_{i},q)\leq r^{2}$).
The fixed radius search is simplest to describe. At each non-terminal
node the query vector is compared to the cut plane's value in the
cut dimension, which chooses one child node as the {}``closer''
and the other as {}``farther''. The tree is recursively descended
depth-first, searching all closer nodes unconditionally. At the terminal
nodes (buckets), the distances from the points of in terminal node
to the query vector are computed explicitly, and any point with distance
less than $r^{2}$ is added to the list of results. As the terminal
node search is a significant performance bottleneck it was implemented
carefully to avoid needless computation and cache misses, and minimize
memory bandwidth.

The farther nodes are also searched, but only if the bounding box
of the farther node intersects the hypersphere of radius $r^{2}$
centered at the query vector. Again for efficiency, this is implemented
in two pieces. First, the perpendicular distance from the query vector
to the cut plane value is computed (as this is available at the node
itself). If this is larger than $r^{2}$ then the farther node is
excluded automatically. If not, then the distance from the query vector
to the closest point in the bounding box of the farther node is computed.
If this distance is less than $r^{2}$ then the farther node is searched,
otherwise it is excluded as well. Theoretically, the perpendicular
test is not necessary as if it rejects searching the farther node,
then so would have the bounding box test. It takes more computation
to test the bounding box, and it was empirically worthwhile to obviate
some of those computations with the simple perpendicular test. 

Searching for $m$ nearest neighbors is similar, except that the $r^{2}$
value changes dynamically during the search. A list of up to $m$
candidate neighbors is maintained during the above search procedure.
The distance to the $m$th neighbor (i.e. the largest distance on
the current list) is the search ball's size, $r^{2}$. In the terminal
node search any point with distance less than this value must replace
one on the working list, and the new $r^{2}$ recomputed as the maximum
value on the list. In the present implementation, this is done with
a priority queue structure based on binary heaps, allowing replacement
of the maximum value and retrieval of the new maximum in $O(\log m)$
time instead of the usual $O(m)$ time for maintaining an always sorted
sorted list. This efficiency gain is important for larger $m$ searches.
At the beginning of the search, $r^{2}$ is set to $+\infty$ and
any point found is added until there are $m$ on the working list.

By default, points are returned in arbitrary order for both searches,
except that for the fixed $m$ search the point with the \emph{largest}
distance will be first on the list because of the priority queue implementation.
There is a subroutine to sort results in ascending order of distance.
This can be called manually by the user, or, if specified at time
of tree creation, will be performed for all searches on that tree.

For many geometrical time-series algorithms, for example, the False
Nearest Neighbor methods for testing embeddings, the points are ordered
in time corresponding to their index. One often wants to find neighbors
in the set close to other existing points in the data set, but exclude
the reference point (which provides the query vector), and a window
of points close in time often known as the {}``decorrelation window''
$W$. The software offers search routines for this task. For these
searches the query vector is set to the value of the reference index
$i$, and in the terminal node search, any candidate neighbor $x_{j}$
is excluded as a valid result if $|i-j|<W$.

\section{Interface and examples}

The codes also offer the option of defining distances along fewer
dimensions than exist in the matrix of input data, i.e, projecting
the input $d$-dimensional vectors to the first $d'<d$ coordinates.

\subsection{Fortran 95}

\begin{singlespace}
\noindent The Fortran code is in \texttt{kdtree2.f90}, in module \texttt{kdtree2\_module}.
The interfaces ought be self-explanatory: \begin{verbatim}
function kdtree2_create(input_data,dim,sort,rearrange) Result (mr)
    ! create a tree from input_data(1:d,1:N)
    ! if PRESENT(dim), use this as dimensionality. 
    ! if (sort .eqv. .true.) then sort all search results by increasing distance.
    real, target                  :: input_data(:,:)
    integer, intent(in), optional :: dim
    logical, intent(in), optional :: sort
    logical, intent(in), optional :: rearrange
    !
    Type (kdtree2), Pointer :: mr  ! the master record for the tree.

subroutine kdtree2_destroy(tp)
    ! Deallocates all memory for the tree, except input data matrix
    Type (kdtree2), Pointer :: tp

subroutine kdtree2_n_nearest(tp,qv,nn,results)
    ! Find the 'nn' vectors in the tree nearest to 'qv' in euclidean norm
    ! returning their indexes and distances in 'indexes' and 'distances'
    ! arrays already allocated passed to this subroutine.
    type (kdtree2), pointer      :: tp
    real, target, intent (In)    :: qv(:)
    integer, intent (In)         :: nn
    type(kdtree2_result), target :: results(:)

subroutine kdtree2_n_nearest_around_point(tp,idxin,correltime,nn,results)
    ! Find the 'nn' vectors in the tree nearest to point 'idxin',
    ! with correlation window 'correltime', returing results in
    ! results(:), which must be pre-allocated upon entry.
    type (kdtree2), pointer        :: tp
    integer, intent (In)           :: idxin, correltime, nn
    type(kdtree2_result), target   :: results(:)

subroutine kdtree2_r_nearest_around_point(tp,idxin,correltime,r2,nfound,nalloc,results)
    type (kdtree2), pointer      :: tp
    integer, intent (In)         :: idxin, correltime, nalloc
    real, intent(in)             :: r2
    integer, intent(out)         :: nfound
    type(kdtree2_result), target :: results(:)

function kdtree2_r_count(tp,qv,r2) result(nfound)
    ! Count the number of neighbors within square distance 'r2'. 
    type (kdtree2), pointer   :: tp
    real, target, intent (In) :: qv(:)
    real, intent(in)          :: r2
    integer                   :: nfound

function kdtree2_r_count_around_point(tp,idxin,correltime,r2)  result(nfound)
    type (kdtree2), pointer :: tp
    integer, intent (In)    :: correltime, idxin
    real, intent(in)        :: r2
    integer                 :: nfound

subroutine kdtree2_n_nearest_brute_force(tp,qv,nn,results) 
    ! find the 'n' nearest neighbors to 'qv' by exhaustive search.
    ! only use this subroutine for testing, as it is SLOW!  The
    ! whole point of a k-d tree is to avoid doing what this subroutine
    ! does.
    type (kdtree2), pointer :: tp
    real, intent (In)       :: qv(:)
    integer, intent (In)    :: nn
    type(kdtree2_result)    :: results(:) 

subroutine kdtree2_sort_results(nfound,results)
    !  Use after search to sort results(1:nfound) in order of increasing 
    !  distance.
    integer, intent(in)          :: nfound
    type(kdtree2_result), target :: results(:) 
\end{verbatim}

\noindent An example follows.\begin{verbatim}
program kdtree2_example
  use kdtree2_module
  type(kdtree2), pointer            :: tree
  integer                           :: N,d
  real, allocatable                 :: mydata(:,:)
  type(kdtree2_result), allocatable :: results(:)

! user sets d, the dimensionality of the Euclidean space
! and N, the number of points in the set. 

allocate(mydata(d,N))   
! note order, d is first, N second. 

! read in vectors into mydata(j,i) for j=1..d, i=1..N

tree => kdtree2_create(mydata,rearrange=.true.,sort=.true.)
! Create the tree, ask for internally rearranged data for speed,
! and for output sorted by increasing distance from the
! query vector

allocate(results(20))
call kdtree2_n_nearest_around_point(tree,idxin=100,nn=20,correltime=50,results)

! Now the 20 nearest neighbors to mydata(*,100) are in results(:) except
! that points within 50 time units of idxin=50 are not considered as valid neighbors.
!
write (*,*) 'The first 10 near neighbor distances are: ', results(1:10)%dis
write (*,*) 'The first 10 near neighbor indexes   are: ', results(1:10)%idx
\end{verbatim}
\end{singlespace}

\subsection{C++}

\begin{singlespace}
\noindent The interface header is \texttt{kdtree2.hpp} and main code
in \texttt{kdtree2.cpp}. The BOOST (\texttt{www.boost.org}) library
must be installed%
\footnote{On the author's Fedora Core 2 Linux system, this can be done by installing
the \texttt{boost} and \texttt{boost-devel} RPM packages.%
} as should the Standard Template library. Interfaces for important
public routines follow. Note that sorting of results in increasing
distance can by done using STL as \texttt{sort(results.begin(),results.end())}.

\noindent \begin{verbatim}
  //constructor
  kdtree2(kdtree2_array& data_in,bool rearrange_in = true,int dim_in=-1);
  // destructor
  ~kdtree2();
  // set to true to always sort
  bool sort_results;

  void n_nearest(vector<float>& qv, int nn, kdtree2_result_vector& result);
  // search for n nearest to a given query vector 'qv'.
  void n_nearest_around_point(int idxin, int correltime, int nn,
      kdtree2_result_vector& result);
  // search for 'nn' nearest to point [idxin] of the input data, excluding
  // neighbors within correltime 
  void r_nearest(vector<float>& qv, float r2,kdtree2_result_vector& result); 
  // search for all neighbors in ball of size (square Euclidean distance)
  // r2.   Return number of neighbors in 'result.size()', 
  void r_nearest_around_point(int idxin, int correltime, float r2,
      kdtree2_result_vector& result);
  // like 'r_nearest', but around existing point, with decorrelation
  // interval. 
  int r_count(vector<float>& qv, float r2);
  // count number of neighbors within square distance r2.
  int r_count_around_point(int idxin, int correltime, float r2);
  // like r_count, but around an extant point.
\end{verbatim}

\noindent An example:\begin{verbatim}

#include <vector>
#include <boost/multi_array.hpp>

using namespace boost;   
using namespace std;   

#include "kdtree2.hpp"

typedef multi_array<float,2> array2dfloat; 

main() {
  kdtree2               *tree;
  int                   N,d
  array2dfloat          mydata;
  kdtree2_result_vector results;

  // user sets d, dimensionality of Euclidean space and
  // N, number of poitns in the set.
  
   mydata.resize(extents[N][dim]);   
   // get space for a N x dim matrix. 

   // read in vectors into mydata[i][j] for i=0..N-1, and j=0..d-1
   // NOTE:  array is in opposite order from Fortran, and is 0-based
   // not 1-based.   This is natural for C++ just as the other was
   // natural for Fortran. In both cases, vectors are laid out
   //  contiguously in memory.

   // notice, no need to allocate size of results, as that will be
   // handled automatically by the STL.  results has most properties
   // of vector<kdtree2_result>.
  
   tree = new kdtree2(mydata,true); // create the tree, ask to rearrange
   tree->sort_results = true;       // sort all results.

   tree->n_nearest_around_point(100,50,20,results);
   // ask for 20 nearest neighbors around point 100, with correlation window
   // 50, push onto 'results'.
}

\end{verbatim}
\end{singlespace}

\section{performance}

We now compare the performance, in searches/s, between KDTREE2 and
the author's previous version in Fortran.

First, a database of 10,000 points chosen randomly and uniformly in
the 3-d unit hypercube (in main CPU cache) query vector chosen likewise,
searching for nearest $m$ neighbors.

\begin{center}\begin{tabular}{|c||r|r|}
\hline 
$m$&
KDTREE2&
old KDTREE\tabularnewline
\hline
\hline 
1&
415843&
367530\tabularnewline
\hline 
5&
190531&
160281\tabularnewline
\hline 
10&
127779&
99919\tabularnewline
\hline 
25&
65359&
41485\tabularnewline
\hline 
500&
4794&
350\tabularnewline
\hline
\end{tabular} \end{center}

For 200,000 points in 3-d unit hypercube (larger than CPU cache).

\begin{center}\begin{tabular}{|c||r|r|}
\hline 
$m$&
KDTREE2&
old KDTREE\tabularnewline
\hline
\hline 
1&
162751&
70712\tabularnewline
\hline 
5&
82904&
31782\tabularnewline
\hline 
10&
57243&
20508\tabularnewline
\hline 
25&
33738&
11075\tabularnewline
\hline 
500&
3001&
261\tabularnewline
\hline
\end{tabular} \end{center}

For 5,000 points in 8-dimensional unit hypercube:

\begin{center}\begin{tabular}{|c||r|r|}
\hline 
$m$&
KDTREE2&
old KDTREE\tabularnewline
\hline
\hline 
1&
36258&
19657\tabularnewline
\hline 
5&
16876&
7608\tabularnewline
\hline 
10&
11790&
4930\tabularnewline
\hline 
25&
7133&
3259\tabularnewline
\hline 
500&
1497&
188\tabularnewline
\hline
\end{tabular} \end{center}

For 50,000 points in 8-dimensional unit hypercube. For the large data
sets in higher dimensions, the new package shows the largest performance
gain.

\begin{center}\begin{tabular}{|c||r|r|}
\hline 
$m$&
KDTREE2&
old KDTREE\tabularnewline
\hline
\hline 
1&
8940&
2050\tabularnewline
\hline 
5&
4338&
874\tabularnewline
\hline 
10&
3144&
601\tabularnewline
\hline 
25&
2069&
359\tabularnewline
\hline 
500&
396&
49\tabularnewline
\hline
\end{tabular}\end{center}

\section{licensing}

The KDTREE2 software is licensed under the terms of the Academic Free
Software License, included in file \texttt{LICENSE} included with
the software. In addition, users of this software must give appropriate
citation in relevant technical documentation or journal paper to the
author, Matthew B. Kennel, Institute For Nonlinear Science, preferably
via a reference to the www.arxiv.org repository of this document.
This requirement will be deemed to be advisory and not mandatory as
is necessary for the purpose of inclusion of the present software
with any software licensed under the GNU General Public License.

This software is downloadable by anonymous FTP at \texttt{ftp://lyapunov.ucsd.edu/pub/nonlinear/kd\_tree/kdtree2.{[}zip|tar.gz{]}.}


\begin{thebibliography}{moore91}
\bibitem[moore91]{Moore91}Moore 91 
\bibitem{Friedman77}aaa 
\bibitem{Omohundro87}om \end{thebibliography}
\end{document}